\documentstyle[amsmath,11pt]{article}

\date{\empty}

\begin{document}

\title{\bf Magnetic tension and gravitational collapse}

\author{{Christos G. Tsagas\thanks{e-mail address:
tsagas@astro.auth.gr}}\\ {\small Section of Astrophysics, Astronomy
and Mechanics, Department of Physics}\\ {\small Aristotle University
of Thessaloniki, Thessaloniki 54124, Greece}\\ {\small DAMTP, Centre
for Mathematical Sciences, University of Cambridge}\\ {\small
Wilberforce Road, Cambridge CB3 0WA, UK}\\ {\small Department of
Mathematics and Applied Mathematics, University of Cape Town}\\
{\small Rondebosch 7701, South Africa}}

\maketitle

\begin{abstract}
The gravitational collapse of a magnetised medium is investigated by
studying qualitatively the convergence of a timelike family of
non-geodesic worldlines in the presence of a magnetic field.
Focusing on the field's tension we illustrate how the winding of the
magnetic forcelines due to the fluid's rotation assists the
collapse, while shear-like distortions in the distribution of the
field's gradients resist contraction. We also show that the
relativistic coupling between magnetism and geometry, together with
the tension properties of the field, lead to a magneto-curvature
stress that opposes the collapse. This tension stress grows stronger
with increasing curvature distortion, which means that it could
potentially dominate over the gravitational pull of the matter. If
this happens, a converging family of non-geodesic lines can be
prevented from focusing without violating the standard energy
conditions.\\\\ PACS number(s): 04.20.-q, 04.40.-b
\end{abstract}

\section{Introduction}
Magnetic fields are very common in astrophysical environments and
stellar magnetism is a long established and very active branch of
astrophysics. Nevertheless, the study of magnetic, and of
electromagnetic, fields in strong gravity environments is less
developed. Most of the available studies address the possible
gravitational effects on the Maxwell field and relatively few look
into the implications of magnetic fields, in particular, for
gravitational collapse itself. One of the most intriguing results so
far has been obtained by Thorne in his analysis of Melvin's
cylindrical magnetic universe~\cite{M}. There, by developing the
concept of `cylindrical energy', the author reached the conclusion
that `a strong magnetic field along the axis of symmetry may halt
the cylindrical collapse of a finite cylinder before the singularity
is reached'~\cite{Th}. The possible support of the Maxwell field
against the gravitational collapse of massive bounded systems was
also studied in~\cite{AP}. That work led to solutions of the
Einstein-Maxwell system where the gravitational attraction is solely
balanced by magnetic stresses. Studies of contracting charged dust
have also suggested that the fluid may `rebounce', thus preventing
black-hole formation~\cite{N}. It has been argued, on the other
hand, that a collapsing spherically symmetric charged dust will
produce naked singularities due to shell-crossing~\cite{O}. Although
these singularities are considered weak because the curvature
invariants and the tidal forces remain finite~\cite{Ne}, their
appearance could also signal that a nonzero Lorentz force and
spherically symmetric charged collapse may not be physically
compatible~\cite{GT}.

In this article we revisit the issue of the magnetic impact on
gravitational collapse from an apparently entirely different
viewpoint. We consider the non-spherical (but not necessarily
cylindrical) collapse of a magnetised fluid, by studying the
convergence of two neighbouring particle worldlines. As it turns
out, we arrive at the same qualitative result as Thorne did, namely
that the gravitational collapse of a magnetised fluid may stop
before reaching the singularity. In our study, however, the reasons
are seemingly unrelated to the cylindrical energy of Melvin's
universe or to the charge density of the magnetised matter. We find
instead that it is the intricate coupling between magnetism and
geometry and the tension properties of the magnetic forcelines,
namely their elasticity, which gives rise to ever increasing
resisting stresses that may prevent the ultimate collapse from
happening. Although these magneto-geometrical tension stresses are
nothing more than the relativistic generalisation of a rather well
known Newtonian effect, their existence remains largely unknown in
the literature. Among the effects of the tension stresses we
identify what is usually referred to as `magnetic braking' and
demonstrate how it assists the collapse of the magnetised fluid. We
also find that shear-like distortions in the distribution of the
field's gradients resist contraction.

Studying the gravitational collapse of a magnetised fluid in full
means solving the complete Einstein-Maxwell equations; a formidable
task even for the most powerful numerical codes~\cite{NOK,DLSS}.
Given that, all analytical studies so far have allowed for a
considerable degree of mathematical simplification. Here we will
consider the question of magnetised collapse without attempting to
solve the Einstein-Maxwell equations. The main aim of this
qualitative study is to shed light upon the role of the magnetic
tension and draw attention to its potential implications. We assume
a highly conductive fluid, similarly to the recent numerical work
of~\cite{DLSSS}, but no axial symmetry. This difference could be the
reason the aforementioned numerical work did not encounter the
resisting magneto-curvature effects discussed here. On the other
hand, it could also be that these tension stresses dominate the
dynamics of the collapse under special circumstances only.

\section{The worldlines of magnetised matter}
We will study the magnetic implications for gravitational collapse
qualitatively by testing the convergence of the particle worldlines.
In doing so we will be using the covariant approach to general
relativity~\cite{E}, and in many respects our discussion will
resemble that of~\cite{TPM}. Throughout this article we assume
conventional matter with positive gravitational mass and pressure,
which means that the standard energy conditions are always
fulfilled. High conductivity means that there is no electric field
and that the magnetic field is `frozen in' with the matter. This is
the well known MHD approximation~\cite{P}.

When looking into the dynamics of gravitational collapse
Raychaudhuri's formula is the key equation, as it covariantly
describes the volume evolution of a self-gravitating fluid
element~\cite{R}. Consider a congruence of timelike worldlines
tangent to the 4-velocity field $u_a$ (with $u_au^a=-1$). These are
the worldlines of the fundamental observers and follow the motion of
the fluid. The Raychaudhuri equation determines the proper-time
evolution of $\Theta=\nabla_au^a$, the scalar measuring the average
contraction (or expansion) between two neighbouring
worldlines~\cite{E}. In a magnetised environment we have~\cite{TB}
\begin{equation}
\dot{\Theta}= -{\textstyle{1\over3}}\Theta^2-
{\textstyle{1\over2}}\kappa\left(\mu+3p+B^2\right)- 2\sigma^2+
2\omega^2+ {\rm D}^a\dot{u}_a+ \dot{u}_a\dot{u}^a\,,  \label{Ray}
\end{equation}
where $\kappa=8\pi G$, $\mu$ and $p$ are respectively the energy
density and pressure of the fluid, $B^2=B_aB^a$ measures the energy
density and the isotropic pressure of the magnetic field ($B_a$),
$\sigma^2$ and $\omega^2$ are the respective magnitudes of the shear
and the vorticity associated with $u_a$ and
$\dot{u}_a=u^b\nabla_bu_a$ is the 4-acceleration. The latter
satisfies the momentum-density conservation law, which for a highly
conductive perfect fluid takes the form~\cite{TB}
\begin{equation}
\left(\mu+p+{\textstyle{2\over3}}B^2\right)\dot{u}_a=-{\rm D}_ap-
\epsilon_{abc}B^b{\rm curl}B^c- \Pi_{ab}\dot{u}^b\,,  \label{Euler}
\end{equation}
where ${\rm D}_a=h_a{}^b\nabla_a$ is the covariant derivative
operator orthogonal to $u_a$ and $\Pi_{ab}=-B_{\langle
a}B_{b\rangle}$ describes the magnetic anisotropic
pressure~\footnote{Angled brackets indicate the symmetric,
trace-free part of projected second-rank tensors (e.g.~$B_{\langle
a}B_{b\rangle}=B_aB_b-(B^2/3)h_{ab}$, where $h_{ab}=g_{ab}+u_au_b$
with $h_{ab}u^b=0$).}. Note that we consider non-geodesic
worldlines, since the motion of the particles is dictated by the
combined Einstein-Maxwell field and not by gravity alone. Also, the
fluid flow is not hypersurface orthogonal which explains the
presence of the vorticity term in (\ref{Ray}).

The right-hand side of Eq.~(\ref{Ray}) determines the dynamics of
the average volume evolution. Terms that are positive definite lead
to expansion, while negative definite terms cause contraction. Thus,
when the standard energy conditions are satisfied, all the
right-hand side terms have a clear-cut role with the exception of
${\rm D}_a\dot{u}^a$, which in principle can go either way. For the
rest of this study, we will focus on ${\rm D}_a\dot{u}^a$ and
examine its potential implications for the final fate of a
collapsing magnetised fluid element.

\section{Lorentz force and magnetic tension}
The magnetic contribution to ${\rm D}_a\dot{u}^a$  comes from the
Lorentz force. The latter is always normal to the direction of the
field lines and emerges whenever the magnetic pattern is distorted
from the condition of local equilibrium. Here, the Lorentz force is
determined by the acceleration vector
\begin{equation}
a_a= -\epsilon_{abc}B^b{\rm curl}B^c= -{\textstyle{1\over2}}{\rm
D}_aB^2+ B^b{\rm D}_bB_a\,,  \label{Lfacc}
\end{equation}
with $a_aB^a=0$. The first gradient in the right-hand side of the
above is due to the magnetic pressure and the second comes from the
field's tension (e.g.~see~\cite{P}). Insofar the tension stress is
not balanced by the gradients of the field's pressure, a net force
is exerted on the fluid particles. Note that $\epsilon_{abc}$ is the
totally antisymmetric alternating tensor orthogonal to $u_a$ and
therefore $a_au^a=0$. Using the 3-Ricci identity (see
Eq.~(\ref{3Ricci}) below), the projected gradient of $a_a$
decomposes as
\begin{equation}
{\rm D}_ba_a= -{\textstyle{1\over2}}{\rm D}_b{\rm D}_aB^2+ {\rm
D}_bB^c{\rm D}_cB_a+ B^c{\rm D}_c{\rm D}_bB_a+ {\cal
R}_{acbd}B^cB^d- 2\omega_{bc}\dot{B}_{\langle a\rangle}B^c\,.
\label{rLf}
\end{equation}
Here ${\cal R}_{abcd}$ is the Riemann tensor of the observer's
instantaneous 3-dimensional rest space and $\omega_{ab}$ is the
vorticity tensor associated with the fluid flow. The last four terms
on the right-hand side of the above convey the magnetic tension
effects. In particular, the magneto-curvature stress in
Eq.~(\ref{rLf}) reflects the special status of vectors, as opposed
to that of scalars, in general relativity. This special status stems
from the geometrical nature of the theory and it is manifested in
the Ricci identity
\begin{equation}
2\nabla_{[a}\nabla_{b]}B_c=R_{dcba}B^d\,,  \label{Ricci}
\end{equation}
applied here to the magnetic field vector, with $R_{abcd}$ being the
spacetime Riemann tensor. When projected orthogonal to $u_a$, the
above expression leads to what is commonly referred to as the
3-Ricci identity~\cite{E}
\begin{equation}
2{\rm D}_{[a}{\rm D}_{b]}B_c={\cal R}_{dcba}B^d-
2\omega_{ab}h_c{}^d\dot{B}_d\,.  \label{3Ricci}
\end{equation}
The Ricci identities argue for a direct interaction between vector
sources and spacetime curvature, which adds to the standard interlay
between matter and geometry as we know it from the Einstein field
equations. In the magnetic case this direct coupling also brings
into play the tension properties of the latter, namely the
elasticity of the magnetic force lines, and couples it in an
intricate way with the geometry of the space. This unique feature,
the magnetic tension, manifests the well known fact that the field
lines `do not like to bend' and react to any attempt that distorts
them. Indeed, within the Newtonian theory the magnetic tension is
known to trigger restoring stresses which depend on the strength of
the field and on the deformation of the magnetic force lines
(measured by their curvature radius)~\cite{P}. What the general
relativistic expression (\ref{rLf}) shows, is that deviations from
Euclidean geometry will lead to analogous tension stresses, which
are also proportional to the magnetic strength and to the amount of
curvature distortion. This time, however, it is the curvature of the
space itself that causes part of the magnetic deformation. In other
words, the magnetic field `feels' the curvature of the space in a
way that is dictated by the Ricci identities and this is
demonstrated by the magneto-curvature stress in the right-hand side
of Eq.~(\ref{rLf}). The effects of this tension stress are generally
counter-intuitive because of the nature of the magnetic property
itself and its subtle coupling with the geometry of the space. The
latter means that even weak magnetic fields can lead to a strong
overall effect under favourable circumstances. The potentially
pivotal implications of magnetism, and of the magnetic tension in
particular, for cosmology were originally discussed in~\cite{MT}.
These results have been confirmed and extended in~\cite{dFSZ},
although an oversight prevented the authors of the latter paper from
recognising the key role of the magnetic tension. Here, we will
consider the implications of the same tension stresses for the
gravitational collapse of a magnetised fluid.

Suppose that both the fluid and the magnetic field have a nearly
homogeneous energy density distribution. In practise this means that
any inhomogeneities that might be present are relatively small and
that we may ignore spatial gradients in the energy density of the
two sources. For the magnetic field our assumption implies that the
Lorentz force term in the right-hand side of (\ref{Euler}) is
dominated by the tension stresses. This is exactly what we need,
since our aim is to investigate the effect of these particular
magnetic stresses. Then, Eq.~(\ref{Euler}) gives $\dot{u}_aB^a=0$
and subsequently it reduces to
\begin{equation}
\left(\mu+p+B^2\right)\dot{u}_a=- \epsilon_{abc}B^b{\rm curl}B^c\,,
\label{Euler1}
\end{equation}
Taking the projected divergence of the above, using the trace of
(\ref{rLf}), with ${\rm D}_aB^a=0$ due to the absence of electric
fields, and substituting into Eq.~(\ref{Ray}) we arrive at
\begin{equation}
\dot{\Theta}+{\textstyle{1\over3}}\Theta^2=-
{\textstyle{1\over2}}\kappa\left(\mu+3p+B^2\right)-
2\left(\sigma^2-\Sigma^2\right)+ 2\left(\omega^2-W^2\right)+
\epsilon^{-1}{\cal R}_{ab}B^aB^b+ \dot{u}_a\dot{u}^a\,, \label{Ray1}
\end{equation}
where $\epsilon=\mu+p+B^2$, $\Sigma^2=({\rm D}_{\langle
a}B_{b\rangle})^2/2\epsilon$, $W^2=({\rm D}_{[a}B_{b]})^2/2\epsilon$
and ${\cal R}_{ab}={\cal R}^c{}_{acb}$ is the 3-Ricci tensor given
by the Gauss-Codacci equation~\cite{E,HE}
\begin{equation}
{\cal R}_{ab}=h_a{}^ch_b{}^dR_{cd}+ R_{acbd}u^cu^d- k_c{}^ck_{ab}+
k_{ac}k^c{}_b\,,  \label{GC}
\end{equation}
where $k_{ab}={\rm D}_bu_a$ is the extrinsic curvature tensor. The
second term in the right-hand side of the above ensures that, in
addition to the usual matter fields, the Weyl curvature (i.e.~tidal
forces and gravitational waves) also contributes to the geometry of
the 3-space. Also note that in deriving Eq.~(\ref{Ray1}) we have
assumed magnetic flux conservation, which guarantees that the
rotation term in the trace of (\ref{rLf}) vanishes.

\section{The magnetic tension stresses}
We will first turn our attention to the quantities $\Sigma^2$ and
$W^2$ in the right-hand side of (\ref{Ray1}), which are the
magnitudes of $\Sigma_{ab}={\rm D}_{\langle
a}B_{b\rangle}/\sqrt{\epsilon}$ and $W_{ab}={\rm
D}_{[a}B_{b]}/\sqrt{\epsilon}$ respectively. The former of these
tensors describes distortions in the distribution of the field
gradients and resists contraction, while the latter is the magnetic
twist tensor and assists the collapse. Note that, although
$\Sigma_{ab}$ and $W_{ab}$ have the shear and the vorticity as their
respective kinematical analogues, their effect on $\Theta$ is
exactly the opposite of the one normally associated with the shear
and the vorticity proper (see also~\cite{PE}). This
counter-intuitive behaviour reflects the fact that both $\Sigma$ and
$W$ carry the tension properties of the field and manifests the
tendency of the magnetic force lines to remain straight. Thus, while
a nonzero shear assists the collapse, the corresponding magnetic
stress tries to balance this effect out. Consider also the tensor
$W_{ab}$, which has magnitude $W^2=W_{ab}W^{ab}/2=({\rm
curl}B_a)^2/4\epsilon$ and is triggered by the winding of the
magnetic field lines around a rotating fluid element. Following
(\ref{Ray1}), a nonzero $W$ will always reduce the gravitational
effect of kinematic vorticity. This manifestation of the field's
tension is also known as `magnetic braking' and can accelerate the
gravitational collapse of a rotating star~\cite{DLSSS}.

For our purposes the key quantity is the magneto-curvature tension
stress ${\cal R}_{ab}B^aB^b$ in the right-hand side of
Eq.~(\ref{Ray1}), which measures the curvature of the 3-space along
the direction of the magnetic force lines. Starting from (\ref{GC})
one can show that, when only the magnetic field is present, ${\cal
R}_{ab}B^aB^b=0$~\cite{T}. Therefore, despite the magnetic energy
input, the curvature of the 3-space in the direction of the field
lines is zero; a result independent of the magnetic strength.
Technically speaking, it is the negative pressure of the field along
its own direction which cancels out the positive contribution of the
magnetic energy density. More intuitively, it is the inherit
tendency of the magnetic force lines to remain `straight' which is
responsible for the aforementioned null result. Clearly, in the
presence of other sources ${\cal R}_{ab}B^aB^b\neq0$. Then, the
magneto-curvature effect on $\Theta$ is rather unexpected. For
${\cal R}_{ab}B^aB^b<0$ the magneto-geometrical stress in
(\ref{Ray1}) brings the particle worldlines closer, but pushes them
apart when the field lines are `positively curved', that is for
${\cal R}_{ab}B^aB^b>0$ (see~\cite{MT} and also~\cite{dFSZ}). This
is against the common perception, which always associates positive
curvature with gravitational contraction. As with the magnetic shear
and the magnetic vorticity stresses discussed earlier, the reason
for the counter-intuitive behaviour of the magneto-curvature term is
the tension properties of the field lines.

Let us assume, mainly for illustration purposes, that the effect of
$\Sigma$ and $W$ is cancelled out by that of their kinematic
counterparts. In other words, we will consider the case where
$\omega^2+\Sigma^2=W^2+\sigma^2$. Although it may not appear so
initially, we will later show that this assumption is much less
restrictive that it looks. For the moment we note that, in addition
to counteracting each other, the pairs $\omega^2$, $W^2$ and
$\sigma^2$, $\Sigma^2$  are of the same nature and `perturbative
order' (i.e.~quadratic in ${\rm D}_au_b$ and ${\rm D}_aB_b$). The
former of these properties supports the assumption that the
aforementioned opposing pairs are very likely to balance each other
out. The latter ensures that these terms, unlike ${\cal
R}_{ab}B^aB^b$ for example, become appreciable only in highly
inhomogeneous configurations. Under such conditions, the
Raychaudhuri equation reads
\begin{eqnarray}
\dot{\Theta}+ {\textstyle{1\over3}}\Theta^2=-R_{ab}u^au^b+ c_{\rm
a}^2\,{\cal R}_{ab}n^an^b+ \dot{u}_a\dot{u}^a\,,  \label{colRay}
\end{eqnarray}
where $n_a=B_a/\sqrt{B^2}$, $c_{\rm a}^2=B^2/\epsilon$ is the
Alfv\'en speed and $R_{ab}$ is the Ricci tensor of the spacetime
with $R_{ab}u^au^b=\kappa(\mu+3p+B^2)/2>0$. The latter means that
the strong energy condition is satisfied.

It should be noted that, in addition to the aforementioned arguments
regarding the relative gravitational input of the shear, the
vorticity and of their magnetic counterparts, bypassing these terms
also helps to isolate the two resisting stresses in
Eq.~(\ref{Ray1}), namely $c_{\rm a}^2{\cal R}_{ab}{\rm n}^a{\rm
n}^b$ and $\dot{u}_a\dot{u}^a$. The latter has been analysed within
a collapsing perturbed Tolman-Bondi spacetime in~\cite{GT}, where it
was found to grow faster than all the other terms in the right-hand
side of the Raychaudhuri equation. Here we will focus on the
magneto-curvature tension stress instead. In any physically
realistic scenario of stellar collapse this term is always positive
and therefore it always resists against further contraction.
Moreover, the strength  of the gravito-magnetic stress is
essentially proportional to the amount of the curvature distortion.
This feature distinguishes the tension stress from the rest and
makes it a very promising candidate for outbalancing the
gravitational pull of the matter.

\section{Gravitational pull vs gravito-magnetic tension}
The assumption of a spatially homogeneous energy density
distribution for the sources means that ${\rm D}_ap=0$, which in
turn guarantees that the 4-acceleration vector $\dot{u}_a$ depends
entirely on the magnetic field (see Eq.~(\ref{Euler1})). Therefore,
when the field is absent all the positive definite terms in the
right-hand side of (\ref{colRay}) vanish and an initially converging
congruence (i.e.~one with $\Theta_0<0$) will focus
(i.e.~$\Theta\rightarrow-\infty$) within a finite amount of time,
unless the energy conditions are violated. This is a fundamental and
well known result about gravitational collapse (e.g.~see~\cite{HE}).
In the magnetic presence, however, the two positive definite terms
in the right-hand side of (\ref{colRay}) will resist the collapse.
Thus, ignoring the supporting effect of $\dot{u}_a\dot{u}^a$, we
argue that the magneto-curvature effects can prevent a converging
congruence of non-geodesic worldlines from focusing, without
violating the standard energy conditions, if
\begin{equation}
c_{\rm a}^2\,{\cal R}_{ab}n^an^b\geq R_{ab}u^au^b\,.  \label{lemma}
\end{equation}
Obviously, the above also holds when the earlier imposed constraint
$\omega^2+\Sigma^2=W^2+\sigma^2$ is replaced by
$\omega^2+\Sigma^2\geq W^2+\sigma^2$. It still holds when
$\omega^2+\Sigma^2<W^2+\sigma^2$, provided that
$\dot{u}_a\dot{u}^a/2>W^2+\sigma^2-\omega^2-\Sigma^2$. So, condition
(\ref{lemma}) holds for a variety of combinations between
$\omega^2$, $\sigma^2$, $W^2$ and $\Sigma^2$. This means that the
earlier restriction placed on these quantities is not essential for
the validity of our argument. The same can also be said about the
fluid inhomogeneities, given that pressure gradients resist the
collapse (through the last term in the right-hand side of
(\ref{colRay})). Our approximations, however, have helped to isolate
and demonstrate the role of the curvature terms in Eq.~(\ref{Ray1}),
which should decide the ultimate fate of the collapse. Indeed, the
main reason for focusing on the two geometrical quantities in the
right-hand side of (\ref{Ray1}) is that, as the collapse proceeds,
we expect the curvature to dominate. Then, the fate of the
collapsing magnetised fluid should be decided by the balance between
the two quantities in Eq.~(\ref{lemma}) and the curvature terms in
(\ref{Ray1}) have not been unduly favoured at the expense of the
rest. If the gravitational pull of the matter, which is the driving
force behind the collapse, is outbalanced, there is a realistic
possibility of avoiding worldline focusing. The magneto-curvature
tension stresses open this possibility. This is so because, while
only the usual matter fields contribute to $R_{ab}u^au^b$, the
tension stress ${\cal R}_{ab}n^an^b$ has additional contributions
from other sources (see Eq.~(\ref{GC})). Probably the most important
among these extra sources is the Weyl curvature. This monitors the
long-range gravitational field and contributes to ${\cal R}_{ab}$
directly via the electric Weyl component~\cite{E}. The latter
describes the tidal forces which are expected to increase
dramatically as the collapse proceeds. These additional
contributions to the 3-Ricci curvature mean that, in principle, the
magneto-curvature tension stresses can outbalance the gravitational
pull of the matter. Although the final outcome depends on the
precise characteristics of the collapse, our analysis points towards
the conjectural but very intriguing possibility that {\em if the
magnetic line deformation due spatial curvature distortions is
strong enough, the resulting tension stresses may just be able to
avert the formation of caustics and eventually the ultimate collapse
of the magnetised fluid.}

\section{Discussion}
The immediate implication of the above study is that violating the
standard energy conditions to prevent an initially converging
congruence from focusing may not be always necessary when magnetic
fields are present. In particular, nonspherical magnetised collapse
(the magnetic presence will inevitably distort spherical symmetry to
a larger or lesser degree) may not end up in a
$\Theta\rightarrow-\infty$ singularity because of the field's
tension. Whether our analysis applies to physical situations, such
as the gravitational implosion of a massive star, depends on what
the exact properties of a realistic collapse are. It might be, for
example, that stellar magnetic fields do not survive to the later
stages of the collapse, or that the MHD limit is no longer a good
approximation. We have no reasons to believe that either of these
possibilities may be true however. In fact, the opposite appears
more likely given the presence of very strong magnetic fields in
compact stellar objects like neutron stars. It might also be that
the gravitational pull of the matter always prevails and condition
(\ref{lemma}) is never satisfied. Very recent numerical simulations
of magnetised hypermassive neutron-star collapse indicate that, at
least when axial symmetry holds, this seems to be the
case~\cite{DLSSS}. If condition (\ref{lemma}) is met, however, the
magnetic tension could stop the convergence of the fluid worldlines
and the same could also happen to the contraction of a magnetised
star. Similarly, if one evolves a magnetised universe backward in
time, they may find a highly curved state that could potentially
re-expand into the past. It should be noted, however, that avoiding
worldline convergence only means avoiding a singularity in the
congruence and does not guarantee a singularity-free spacetime; at
least under the current consensus of what a singularity
is~\cite{HE}. For example, the spacetime can still be geodesically
incomplete. Nevertheless, any such singularities should be of
limited influence if most of the matter can successfully avoid them.

Stresses that support against gravitational collapse are not
exclusively particular to a magnetic presence. It is well known that
whenever the worldlines are not hypersurface orthogonal or
geodesics, supporting stresses always appear due to rotation or
pressure gradients. Vorticity, for example, has been considered in
the past as a possible way of preventing the ultimate
collapse~\cite{Ma}. Therefore, it is not so much the existence of
magnetic related stresses that resist gravitational collapse, as the
nature of the stresses themselves. In this respect, our key result
is that, when magnetic fields are involved, one of the supporting
stresses depends (in fact it is proportional to) on the distortion
of the curvature itself. The presence of this stress is a direct and
inevitable consequence of the vector nature of magnetic fields and
of the geometrical nature of general relativity, while their
counter-intuitive effects result from the tension properties of the
field. It is the elasticity of the magnetic force lines, their
inherit tendency to remain `straight', which manifests itself as a
reaction to curvature distortions that is proportional to the
distortion itself. In a sense, it appears as though the elastic
properties of the field have been injected in to the fabric of the
space.

Finally, when considering the unconventional magnetic behaviour
described so far, the reader should keep in mind that magnetic
fields are rather unusual sources themselves. The Maxwell field is
the only vector source that we know that exists in the universe
today. Within the geometrical framework of general relativity, the
status of vector fields is different from that of the ordinary
matter. This is so because, in addition to the Einstein field
equations, vector sources interact with the spacetime geometry
through the Ricci identities as well. This purely geometrical
coupling, which has been at the centre of our discussion, is already
known to trigger some other rather nontrivial effects. The best
known example is probably the `scattering' of electromagnetic
radiation by the gravitational field, which leads to what is
commonly referred to in the literature as `wave tales'~\cite{dWB}.
Here, we have outlined the potential implications of effectively the
same Einstein-Maxwell coupling for gravitational collapse.

\section*{Acknowledgements}
The author wishes to thank Mihalis Dafermos, George Ellis, Cristiano
Germani, Kostas Kokkotas, Leon Mestel, Nikos Stergioulas and John
Stewart for helpful discussions and comments.

\end{document}